\documentclass[]{spie}  %>>> use for US letter paper
\usepackage{amsmath,amsfonts,amssymb}
\usepackage{graphicx}
\usepackage[colorlinks=true, allcolors=blue]{hyperref}

\usepackage{changes}
\usepackage{lipsum}% <- For dummy text
\definechangesauthor[name={Xinyue Wang}, color=red]{Xinyue}

\title{Iterative-in-Iterative Super-Resolution Biomedical Imaging Using One Real Image}

\author[a]{Yuanzheng Ma}
\author[a]{Xinyue Wang}
\author[b]{Benqi Zhao}
\author[b]{Ying Xiao}
\author[c]{Shijie Deng}
\author[a]{Jian Song}
\author[a,*]{Xun Guan}

\affil[a]{Institute of Data and Information, Tsinghua Shenzhen International Graduate School, Tsinghua University, Shenzhen, 518055, China}
\affil[b]{Bejiing Tsinghua Changgung Hospital, School of Clinical Medicine, Tsinghua University, Beijing, 102218, China}
\affil[c]{School of Optoelectronic Engineering, Guilin University of Electronic Technology, Guilin, 541004, China}

\authorinfo{Further author information: (Send correspondence to Xun Guan)\\Xun Guan: E-mail: xun.guan@sz.tsinghua.edu.cn}

\begin{document} 
\maketitle

\begin{abstract}
Deep learning-based super-resolution models have the potential to revolutionize biomedical imaging and diagnoses by effectively tackling various challenges associated with early detection, personalized medicine, and clinical automation. However, the requirement of an extensive collection of high-resolution images presents limitations for widespread adoption in clinical practice. In our experiment, we proposed an approach to effectively train the deep learning-based super-resolution models using only one real image by leveraging self-generated high-resolution images. We employed a mixed metric of image screening to automatically select images with a distribution similar to ground truth, creating an incrementally curated training data set that encourages the model to generate improved images over time. After five training iterations, the proposed deep learning-based super-resolution model experienced a 7.5\% and 5.49\% improvement in structural similarity and peak-signal-to-noise ratio, respectively. Significantly, the model consistently produces visually enhanced results for training, improving its performance while preserving the characteristics of original biomedical images. These findings indicate a potential way to train a deep neural network in a self-revolution manner independent of real-world human data.
\end{abstract}

% Include a list of keywords after the abstract 

\section{INTRODUCTION}
\label{sec:intro}  % \label{} allows reference to this section

Biomedical imaging plays a significant role in early detection, intraoperative monitoring, and follow-up care \cite{qu2008imaging, razzak2018deep, fu2023field}. However, collecting high-resolution biomedical images involves complex setups and post-processing, posing a trade-off between temporal and spatial resolution \cite{xu2017enhanced, wang2023hybrid}. Moreover, high-resolution biomedical images are frequently associated with high-intensity laser and density sampling, leading to potentially irreversible damage to organisms \cite{zhao2021deep}. For instance, acquiring a high-resolution angiography image through scanning-based photoacoustic imaging can take up to 10 minutes \cite{zhou2019optical}, and traditional Magnetic Resonance Imaging (MRI) requires intensive post-processing to obtain accurate organism images \cite{behrenbruch2004image, takagi2023high}.

Some researchers have proposed the restoration of resolution through deep neural networks, which would allow the focus to shift towards improving imaging speed during data collection \cite{li2021review, ma2022cascade}. However, training such models necessitates an extensive number of high-resolution images, which poses a challenge in the context of limited biomedical datasets for deep learning. The shortage of biomedical data raises the question of whether it is possible to train a neural network using a minimal number of high-resolution images or even just one. If feasible, this approach would not only alleviate the need for reducing the density of high-intensity laser usage to protect organisms but also address the fundamental issue of insufficient biomedical datasets for deep learning models.

Previous attempts to address the issue of insufficient training images include transfer learning methods, where images from other domains are used for pre-training, followed by refining the weights using a small number of images from the target domain \cite{li2021review}. However, biomedical images often exhibit substantial differences in imaging modalities, resolutions, and acquisition protocols. The huge bias between different data domains limit the effectiveness of transfer learning, as the pre-training data may not adequately capture the specific characteristics of biomedical images \cite{shin2016deep}. Some researchers have proposed using regularization techniques to train the model to prevent overfitting and improve model generalization \cite{farsiu2004fast}. However, these methods have limitations in achieving significant improvements, and techniques such as weight decay and batch normalization may result in overly smooth outputs without preserving fine texture \cite{li2021review}. Another approach, which is data augmentation, requires the manual increase of the size of the training dataset through random rotations, flips, translations, and other transformations \cite{shorten2019survey}. While traditional data augmentation techniques like random cropping, resizing, or color jittering are effective for object recognition tasks, they may not accurately simulate the complex degradation processes in super-resolution imaging. Consequently, the augmented samples may not fully represent the real-world variations encountered during super-resolution tasks.

In this paper, we propose an approach where the super-resolution diffusion model is employed to generate images that are subsequently used for self-training. By progressively incorporating more similar images generated by the model, we effectively enhance its performance in external iterations of the training process. Remarkably, we demonstrate the successful reconstruction of undersampled biomedical images using only a single high-resolution image to train the super-resolution model. The results reveal a progressive improvement in structure similarity and peak signal-to-noise ratio, increasing from 0.51114 to 0.563 and from 22.75 to 24.67, respectively. Moreover, as the number of external iterations increases, we observe a significant enhancement in the diversity of the generated images, which can be utilized to train the model itself. These findings indicate that the neural network can generate novel images based on limited input data and utilize them for further self-training. The associated code is available on GitHub: https://github.com/yuanzhengthu/Iterative-in-Iterative-SR-for-Biomedical-Imaging-Using-a-Single-HR-Authentic-Image.
 
\section{METHODS}
\subsection{Mixed metric for screening high-quality generated images}
The Fr\'echet Inception Distance (FID), based on feature representations extracted from a pre-trained Inception-V3 network, is commonly used as a quantitative measure for assessing the performance of image generation models\cite{shin2021deep}. The function for evaluating the image with FID is as follows:
\[
\text{FID} = \lVert \mu_{\text{real}} - \mu_{\text{fake}} \rVert^2 + \text{Tr}(\Sigma_{\text{real}} + \Sigma_{\text{fake}} - 2 \sqrt{\Sigma_{\text{real}} \Sigma_{\text{fake}}}),
\]
where the $\mu_{real}$ and $\mu_{fake}$ are mean vectors of feature representations of real and generated images, respectively. And $\sigma_{real}$ and $\sigma_{fake}$ are covariance matrices of feature representations of real and generated images, respectively. The operator $||.||^2$ is squared Euclidean distance, and operator $Tr(.)$ is trace of matrix\cite{shin2021deep}. In our experiments, initially, we utilized only FID to assess the quality of generated images by comparing them to real samples. However, we encountered challenges due to variations in FID values among different real samples, resulting from inherent biases in the dataset. To address this issue, we introduce the Inception Score (IS) as an additional evaluation metric\cite{borji2022pros}. The IS is specifically designed to be independent of dataset biases, making it well-suited for evaluating the quality of generated images across different scenarios.

The IS can be calculated using the following equation:
\[
\text{IS} = \exp\left(\mathbb{E}\left[\text{KL}(p(y|x) \, || \, p(y))\right]\right),
\]
where the operator $\mathbb{E}\left[.\right]$ represents the expectation over the generated images. The term $\text{KL}(p(y|x) , || , p(y))$ refers to the Kullback-Leibler divergence between two probability distributions. Specifically, $p(y|x)$ represents the class probability distribution obtained from the Inception-V3 model for the generated image $x$, while $p(y)$ denotes the marginal class distribution, typically assumed to be a uniform distribution.

The mixed score of FID and IS can be calculated using the following equation:
\[
\text{M} = (\text{IS} > T1)\ \&\ (\text{FID} < T2)
\].
This equation determines whether a generated image has an IS score greater than T1 and an FID score less than T2 simultaneously. If the generated image satisfies both conditions, it is considered relevant to the real domain, and it is then included in the training dataset for the next external iteration to train the model further.

\subsection{Enhancing the performance of super-resolution models through iterative iterations}
\begin{figure} [ht]
   \begin{center}
   \begin{tabular}{c} %% tabular useful for creating an array of images 
   \includegraphics[width=14cm]{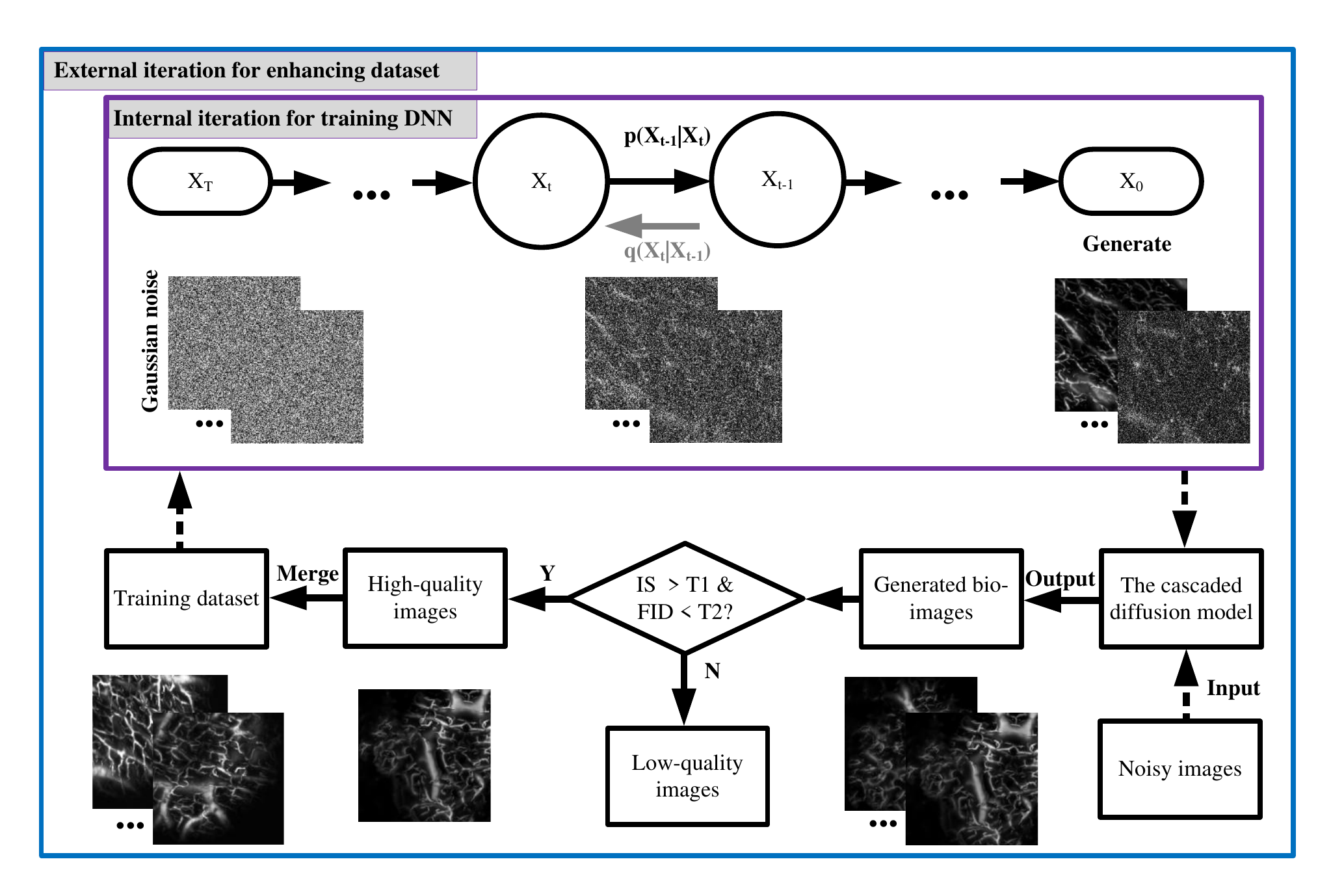}
   \end{tabular}
   \end{center}
   \caption[example] 
%>>>> use \label inside caption to get Fig. number with \ref{}
   { \label{fig1} 
The schematic of the iterative in iterative super-resolution method for biomedical imaging. The internal iterations represent the training of the super-resolution diffusion neural network, while the external iterations focus on enhancing the dataset. The thresholds T1 and T2 are utilized to evaluate the performance using the IS and FID, respectively. DNN: Diffusion neural network. $x_{t}$: the state of the image in the diffusion process.}
   \end{figure}

Given the inherent flexibility of super-resolution models, we embarked on investigating their potential for simultaneous reconstruction and generation. We introduced an external iterative loop outside the training iterations, as illustrated in Fig. \ref{fig1}. At the end of each external iteration, this approach enabled us to generate new realistic images for training, leveraging the existing high-resolution image(s) and progressively enhancing the performance of the diffusion-based model \cite{li2022srdiff,ho2022cascaded}.
% Note: If compiling with LaTeX+dvipdf, please ensure images generated from 
% other software packages have their bounding boxes set correctly.

In our approach, we introduce a carefully designed noise into the training process of the model to alter the initial conditions of the diffusion process\cite{ma2023self}. This allows us to generate new high-resolution images after each training iteration. Subsequently, these generated images undergo screening using the mixed score evaluation method mentioned earlier. High-quality images selected through this screening process are then incorporated into the training dataset, motivating the model to generate even more improved results.

Initially, we set the threshold values as $[T1, T2] = [10, 500]$ in this experiment. However, to further encourage the model to improve its performance, we gradually increase T1 and decrease T2. This adjustment aims to enhance the model's ability to generate superior results with higher fidelity and improved perceptual quality.

\subsection{Noise overlapping from different dimensions enabling diffusion model to generate more biomedical images}
In our previous research, we introduced an approach for generating realistic angiography images by utilizing hand-drawn doodles as a basis for training a super-resolution model. Specifically, our image generation model is based on the diffusion-based super-resolution method\cite{ma2023self}. This method takes advantage of spatial locality, temporal invariance, and statistical stationarity to process low-resolution images using diffusion operators and parameter sharing, thereby producing high-resolution images\cite{ho2020denoising,yue2015beyond,zhang2018unreasonable}.

However, when dealing with biomedical images with more structural information, such as MRI images, we encounter challenges in directly generating specified noise to alter the image generation process. To address this, we introduced randomly selected resize factors to the low-resolution images, treating each variation as a distinct low-resolution image. By doing so, we can effectively generate more realistic super-resolution images of MRI, which can be utilized to progressively enhance the performance of super-resolution.

\section{RESULTS}
\subsection{Angiography images generation with noisy images}
In this experiment, our focus is on training a diffusion model to restore four times undersampled high-resolution angiography images to their original resolution. To fully leverage the underdetermined nature of the super-resolution model, we take it a step further and undersample the already four times undersampled high-resolution angiography images, resulting in 16 times undersampled versions. Subsequently, we utilize either the 16 times or cascaded four times super-resolution model to recover the resolution, yielding a more diverse collection of high-fidelity angiography images\cite{ho2022cascaded}.

\begin{figure} [h]
   \begin{center}
   \begin{tabular}{c} %% tabular useful for creating an array of images 
   \includegraphics[width=14cm]{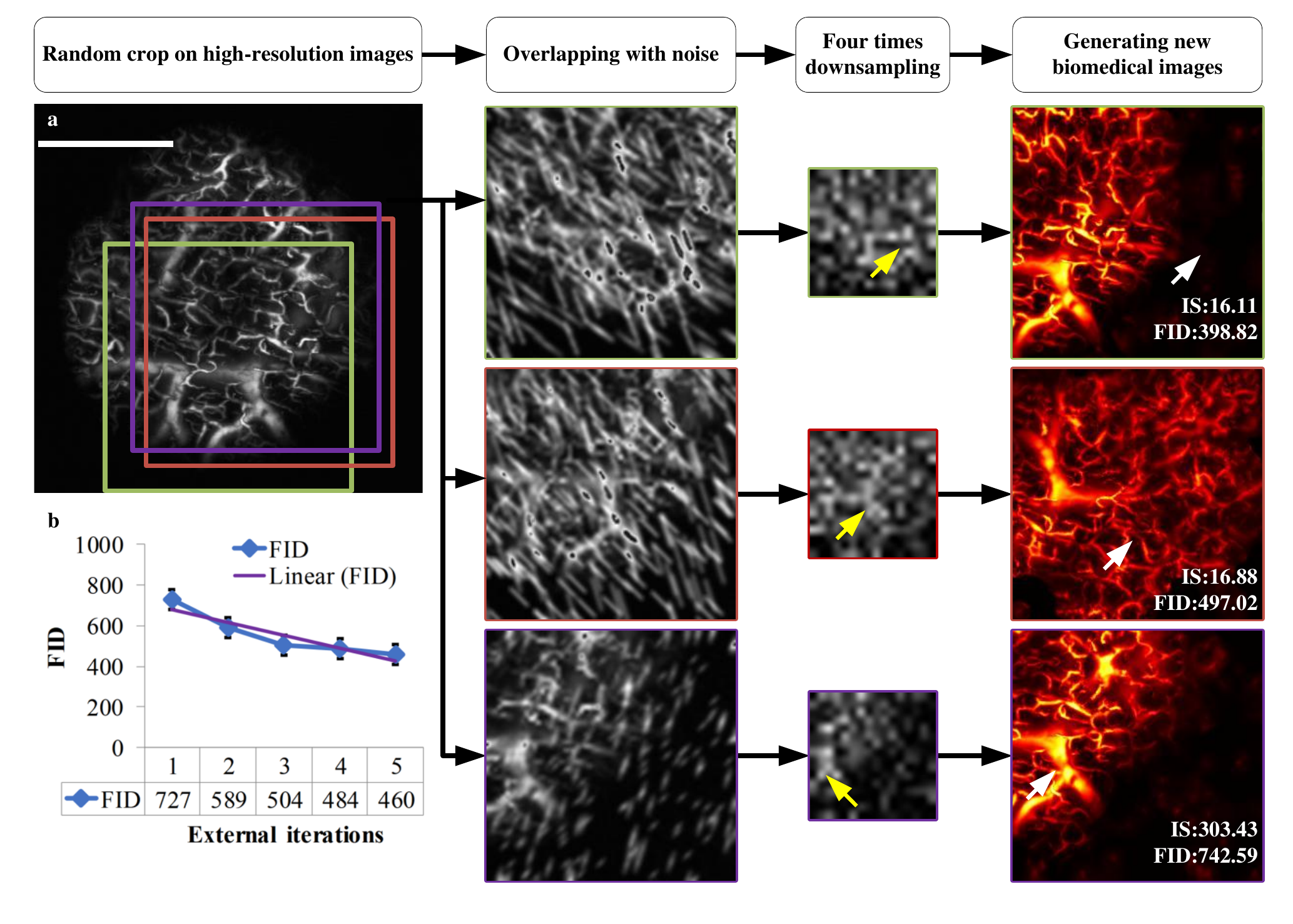}
   \end{tabular}
   \end{center}
   \caption[example] 
%>>>> use \label inside caption to get Fig. number with \ref{}
   { \label{fig2} 
\textbf{a}: The demonstration of randomly sampling from a high-resolution angiography image to generate various high-resolution versions by overlaying different types of noise. \textbf{b}: The maximum FID and IS between the generated images and real samples decreases as the number of external iterations increases. Scale bar: 1mm.}
   \end{figure}

\begin{figure} [h]
   \begin{center}
   \begin{tabular}{c} %% tabular useful for creating an array of images 
   \includegraphics[width=14cm]{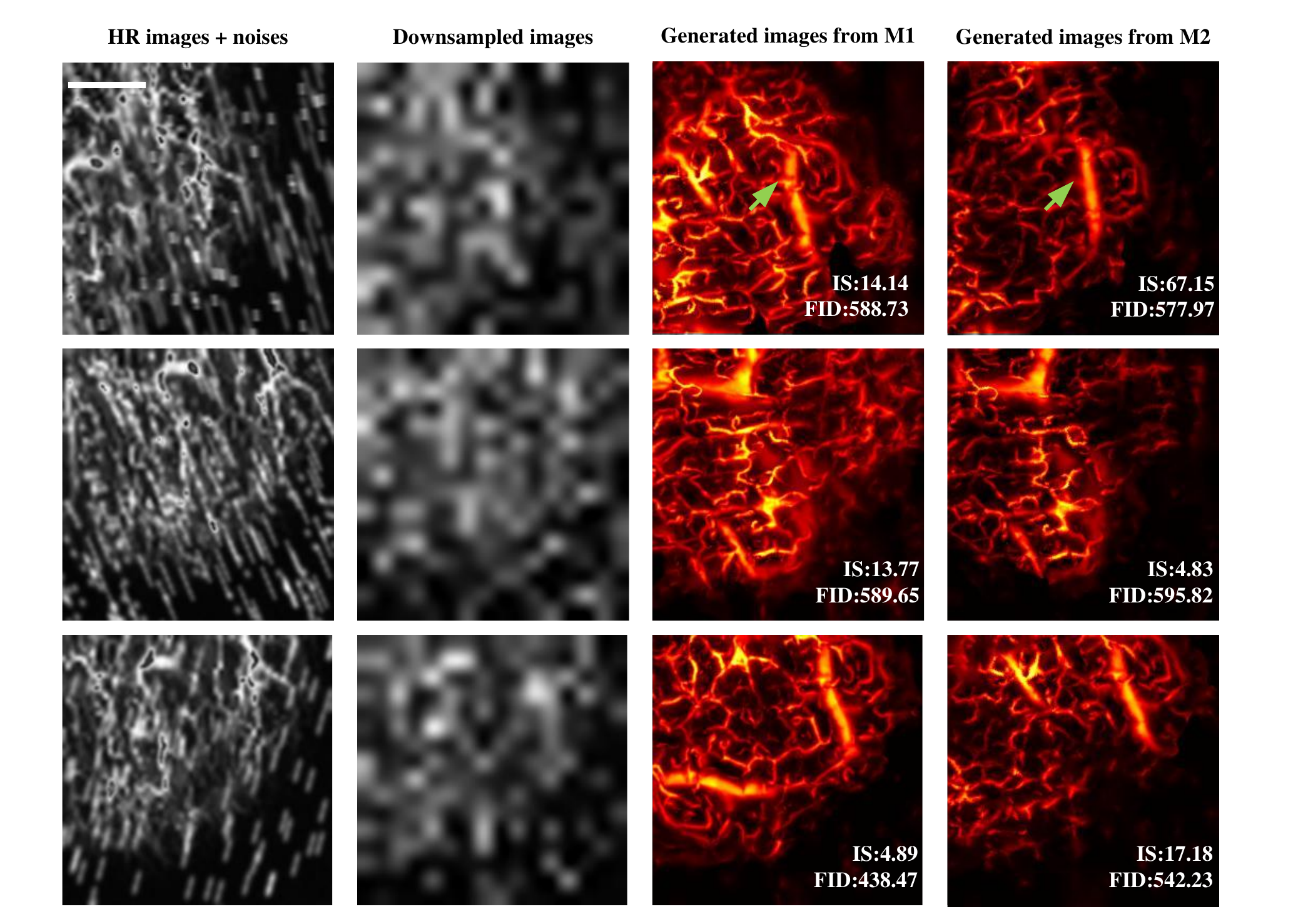}
   \end{tabular}
   \end{center}
   \caption[example] 
%>>>> use \label inside caption to get Fig. number with \ref{}
   { \label{fig3} 
The realistic biomedical images generated by different generation models, namely model 1 (M1) and model 2 (M2). It is important to note that both M1 and M2 were trained using the same single image as input.}
   \end{figure}

As depicted in Fig. \ref{fig2} \textbf{a}, the highlighted areas with arrows demonstrate that different realistic high-resolution images can be generated from the same portion of the original high-resolution image. These variations are achieved through different overlapping noises and a large undersampling ratio, illustrating the impact of these factors on the generated images. In Fig. \ref{fig2} \textbf{b}, it can be observed that with an increase in the number of external iterations from 1 to 5, there is a consistent trend of decreasing the maximum FID between the generated images and real samples. Simultaneously, the maximum IS between the generated images and real samples shows an increasing trend.
To ensure the validity of our approach in the future, we employed various models to process a noise-overlapped image. The outcomes, depicted in Fig. \ref{fig3}, demonstrate the generation of diverse and realistic biomedical images. Notably, the highlighted blood vessel, indicated by the arrows, was produced using the identical noise distribution.
   
\subsection{Under-sampled photoacoustic images reconstruction}
\begin{figure} [ht]
   \begin{center}
   \begin{tabular}{c} %% tabular useful for creating an array of images 
   \includegraphics[width=14cm]{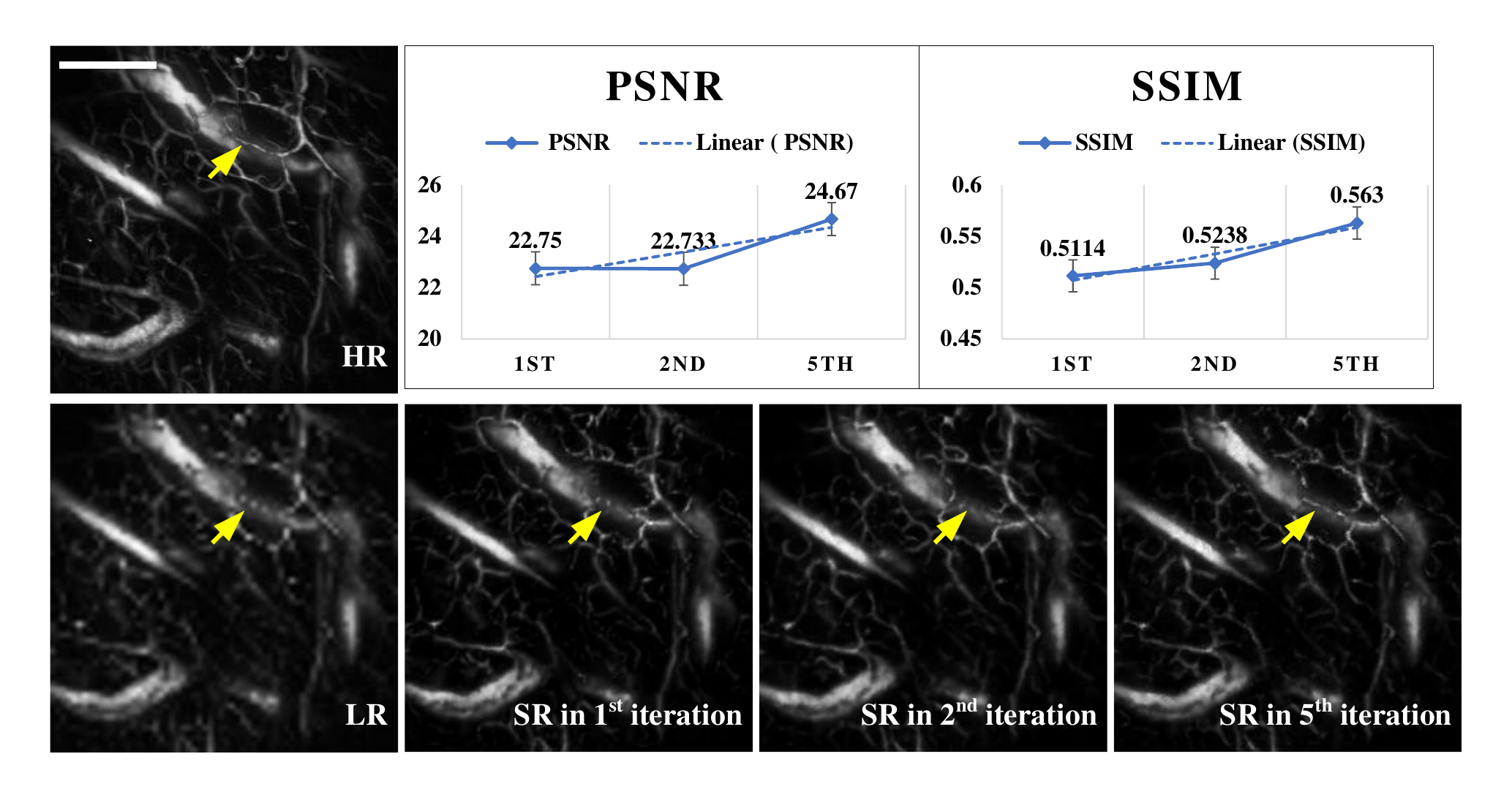}
   \end{tabular}
   \end{center}
   \caption[example] 
%>>>> use \label inside caption to get Fig. number with \ref{}
   { \label{fig4} As the number of external iterations increases, there is a gradual improvement in the reconstruction of angiography images. HR: The high-resolution image. LR: The low-resolution image. SR: The super-resolution image. Scale bar: 0.5mm.}
   \end{figure}

As the models generate increasingly realistic angiography images, they are gradually incorporated into the training dataset, significantly improving the model's ability to reconstruct under-sampled angiography images. This visual enhancement is demonstrated in Fig. \ref{fig4}. 

As depicted in Fig. \ref{fig4}, as the number of external iterations increases, there is a notable improvement in both the Peak Signal-to-Noise Ratio (PSNR) and Structure Similarity (SSIM) metrics. The PSNR increases from 22.75 to 24.67, while the SSIM improves from 0.5114 to 0.563. These quantitative improvements highlight the enhanced image quality and fidelity achieved through the iterative process.

Furthermore, Fig. \ref{fig4} includes yellow arrows highlighting a specific observation. In the 5th iteration, a previously unseen thin blood vessel became visible in the super-resolution image. This visual representation emphasizes the capability of the model to capture fine details and subtle features that were not apparent in the initial iterations, indicating the effectiveness of the proposed approach in enhancing angiography images.
Throughout our experiments, we observed certain cases in which the model did not successfully generate high-resolution angiography images. However, we found that the reconstructed images showed noticeable improvements after several iterations. This observation emphasizes the importance of implementing additional measures to ensure the inclusion of only high-quality images in the training dataset.

\subsection{Incorporating random resizing disturbance for training one-shot super-resolution models of MRI brain images}
\begin{figure} [ht]
   \begin{center}
   \begin{tabular}{c} %% tabular useful for creating an array of images 
   \includegraphics[width=14cm]{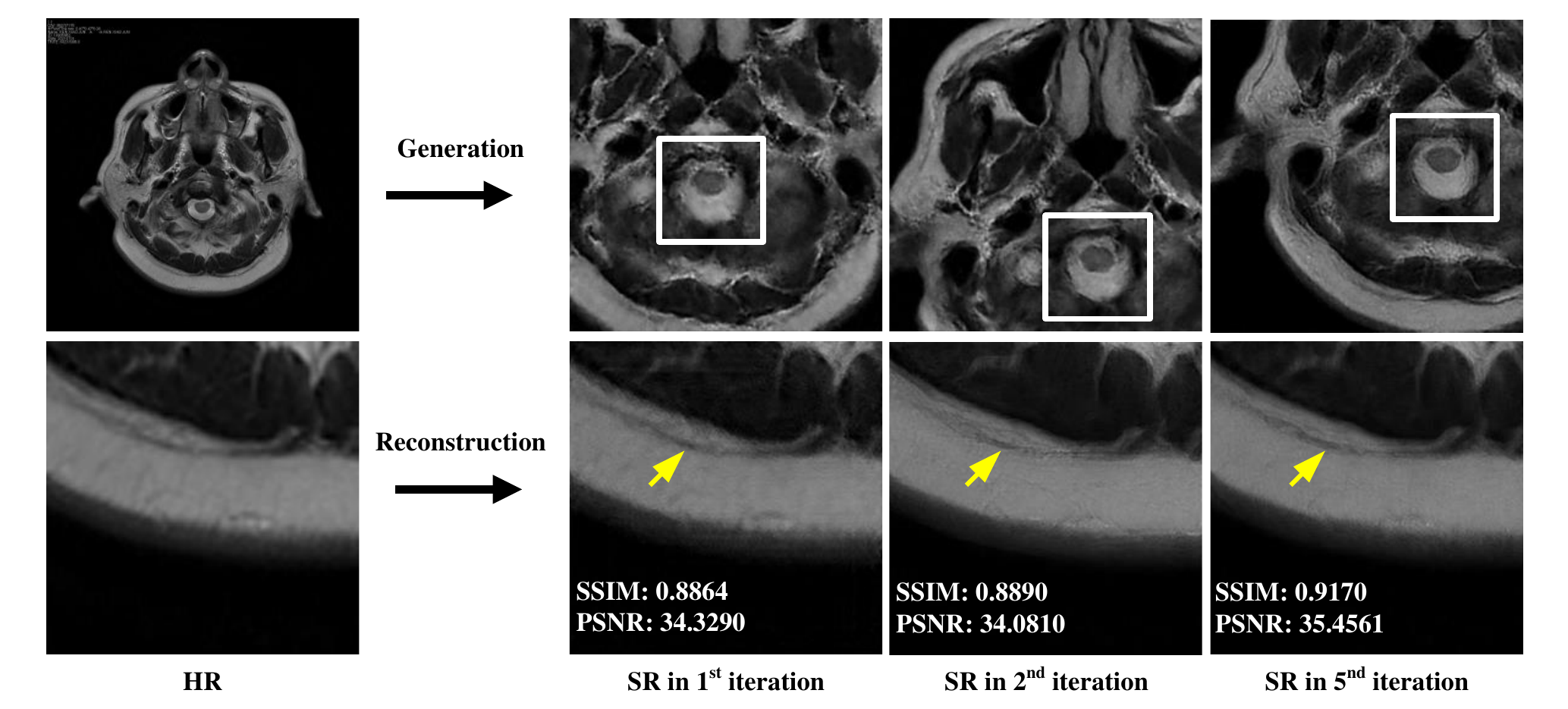}
   \end{tabular}
   \end{center}
   \caption[example] 
%>>>> use \label inside caption to get Fig. number with \ref{}
   { \label{fig5} 
An illustrative demonstration of the capabilities of a super-resolution model trained using only one high-resolution MRI image in image generation and reconstruction. HR: The high-resolution image. SR: The super-resolution image.}
   \end{figure}

To generate a diverse set of high-resolution images for training super-resolution models, we introduced random fluctuations during the resizing process of low-resolution MRI images. The impact of this approach is visually demonstrated in Fig. \ref{fig5}. Remarkably, as the number of iterations increases, the reconstructed images showcase the improved quality and smooth texture, as supported by the observed enhancements in PSNR and SSIM.

As shown in Fig. \ref{fig5}, this visual representation highlights the progressive improvement in image quality achieved through the iterative process. The figure includes yellow arrows that indicate an increase in restored detail as the number of external iterations increases. In addition, Fig. \ref{fig5} incorporates white rectangles to denote the generation of better MRI images corresponding to the increasing number of external iterations.

\section{DISCUSSION AND CONCLUSION}
This study introduces an approach to training super-resolution models for biomedical imaging. By harnessing self-generated high-resolution images and employing a Fréchet Inception Distance-based image screening method, our proposed approach effectively trains models with minimal data, even with a single authentic image. The iterative training process over five generations consistently enhances performance, as evidenced by improvements in structure similarity and peak signal-to-noise ratio compared to the original performance.

A notable advantage of our approach is the continuous improvement observed without reaching a plateau. The model generates visually enhanced output while preserving the original structural similarity. These findings highlight the potential of this self-evolving approach to training deep neural networks independently of real-world human data. This approach holds great promise in addressing challenges in biomedical imaging, such as early detection, personalized medicine, and clinical automation. 

However, it is crucial to emphasize that the proposed method requires a careful design design of the image screening process to guarantee the inclusion of only high-quality images in the dataset. Striking the right balance is essential because if the screening is excessively strict, the model may tend to replicate images from the training dataset. Conversely, if the screening method is not stringent enough, the model may incorporate more low-quality images, which could adversely affect the training process and overall performance. Additionally, we observed that the diffusion-based super-resolution method employed in our approach tends to introduce a certain smoothing effect during the diffusion process, resulting in the loss of fine details and edges in the reconstructed super-resolution images, especially when insufficient images are generated during the iterations.

Further research and validation are necessary to fully explore the potential and applicability of this self-revolutionary training approach in real-world biomedical imaging scenarios, where only one real image is available as the training dataset. Additionally, we plan to conduct additional experiments to determine the saturation point of our method, as there is still room for further progress and improvement. Addressing the challenges related to noise image design and mitigating the potential loss of fine details are crucial areas that warrant thorough investigation in future studies.

\acknowledgments % equivalent to \section*{ACKNOWLEDGMENTS}       
 
This work was supported in part by the National Key R\&D Program of China (2021YFA1001000); in part by Shenzhen Science, Technology and Innovation Commission: Fundamental Research Scheme General Program (JCYJ20220530142809022) and Stable Support (WDZC20220811170401001). 

% References
\bibliography{report} % bibliography data in report.bib
\bibliographystyle{spiebib} % makes bibtex use spiebib.bst

\end{document}